\documentclass[a4paper,11pt]{article}
\usepackage{jcappub}

\usepackage[colorlinks=true,urlcolor=blue,linkcolor=blue]{hyperref}
\usepackage{graphicx}
\usepackage{epsfig}
\relax

\def\be{\begin{equation}}
\def\ee{\end{equation}}
\def\bs{\begin{subequations}}
\def\es{\end{subequations}}

\def\be{\begin{equation}}
\def\ee{\end{equation}}
\def\bs{\begin{subequations}}
\def\es{\end{subequations}}
\newcommand{\een}{\end{subequations}}
\newcommand{\ben}{\begin{subequations}}
\newcommand{\beq}{\begin{eqalignno}}
\newcommand{\eeq}{\end{eqalignno}}

\newcommand{\apj}{ApJ}
\newcommand{\mnras}{MNRAS}
\newcommand{\prd}{PhysRevD}
\newcommand{\jcap}{JCAP}
\newcommand{\aap}{A\&A}
\newcommand{\aj}{AJ}
\newcommand{\apjl}{ApJL}

\newcommand\fverb{\setbox\pippobox=\hbox\bgroup\verb}
\newcommand\fverbdo{\egroup\medskip\noindent%
                        \fbox{\unhbox\pippobox}\ }
\newcommand\fverbit{\egroup\item[\fbox{\unhbox\pippobox}]}
\newbox\pippobox

\title{Where the world stands still: turnaround as a strong test of  $\Lambda$CDM cosmology}

\author[a,b]{V. Pavlidou}
\author[a]{T.~N. Tomaras}
\affiliation[a]{Department of Physics and ITCP\footnote{ Institute of Theoretical and Computational Physics (formerly Institute of
Plasma Physics)}, University of Crete, University of Crete, 71003 Heraklion,Greece} 
\affiliation[b]{Foundation for Research and Technology - Hellas, IESL, Voutes, 71110 Heraklion, Greece} 
\emailAdd{pavlidou@physics.uoc.gr}
\emailAdd{tomaras@physics.uoc.gr}

\abstract{
Our intuitive understanding of cosmic structure formation works best
in scales small enough so that isolated, bound, relaxed gravitating systems are no longer adjusting their radius; and large enough so that space and matter follow the average expansion of the Universe. Yet one of the most robust predictions of  $\Lambda$CDM cosmology concerns the scale that separates these limits: the turnaround radius, which is the non-expanding shell furthest away from the center of a bound structure.    
We show that the maximum possible value of the turnaround radius within the framework of the $\Lambda$CDM model is, for a given mass $M$, equal to $(3GM/\Lambda c^2)^{1/3}$, with $G$ Newton's 
constant and $c$ the speed of light, {\em independently} of cosmic
epoch, exact nature of dark matter, or baryonic effects.  We discuss
the possible use of this prediction as an observational test for
$\Lambda$CDM cosmology. Current data appear to favor $\Lambda$CDM over
alternatives with local inhomogeneities and no $\Lambda$. However
there exist several local-universe structures that have, within errors, reached their
limiting size. With improved determinations of
their turnaround radii and the enclosed mass, these objects may challenge the limit and
$\Lambda$CDM cosmology. }

\begin{document}

\maketitle
\flushbottom

\section{Introduction}
\setcounter{equation}{0}
 In spite of great successes of the $\Lambda$CDM  model in the
 interpretation of cosmological observations, including the cosmic
 microwave background, supernovae type Ia, and the large scale
 structure in the universe, alternative cosmologies continue to
 attract interest and cause intense debate. At the heart of the issue
 lies the -unknown- physics of the dark sector.   

Weakly interacting particle dark matter has by now accumulated
astrophysical evidence that goes beyond its ``relic abundance
miracle'' appeal (e.g. \cite{larsb,gianfranco}). For example, combined imaging of
merging clusters both through the X-ray emission of shocked gas
(tracing the collisional baryonic component) and through lensing
(tracing the gravitational potential peaks), have shown a clear
separation between the two and very different geometries between them
\cite{marusa,marusa2,markevich}. However, the up to now lack of any sign for TeV-scale
supersymmetry or an indisputable signature of any of the prime candidates for particle dark
matter\footnote{despite several potentially
interesting signals (e.g., most recently, the
Galactic Center gamma-ray signal \cite{finkbeiner} and the X-ray signal from galaxy
clusters \cite{bulbul,andromeda})} has intensified the analysis of alternative proposals as
resolutions of the ``dark matter" puzzle, with Modified Newtonian
Dynamics (MOND) being the most popular among them (e.g.,
\cite{famaey,milgrom}), although attempts to reconcile the observations of
merging clusters with MOND require the existence of at least {\em
  some} particle dark matter, such as sterile neutrinos
\cite{angus}.

 Our current best estimates on the other component of the dark sector,
 dark energy \cite{planck}, indicate that it behaves effectively as a
 cosmological constant, but with an uncomfortably low value, defying
 both microscopic and ``anthropic'' interpretations \cite{weinberg}.

Without a confirmed microphysical foundation, $\Lambda$CDM is
therefore our current-best effective theory of the evolution of the
Universe. Observations that might contradict its predictions and
illuminate the direction towards the necessary new physics to complete
our understanding of the cosmos are thus actively pursued. However,
these are frequently hindered by unavoidable complexities in
connecting theory with the real world. For example, there is
persisting tension between observations and predictions of the cold
dark matter (CDM) model for relatively small scales, such as the
abundance of dwarf galaxies and the central profiles of dark matter
halos (e.g. \cite{mondreview}). At these scales though, the (much more
complex and difficult to simulate) behavior of luminous matter plays a
dominant role in determining both the shape of central profiles and
the observability of small satellite galaxies  (e.g.,
\cite{baryons1,baryons2}; see however \cite{baryons3} for
counter-arguments on the efficacy of baryonic physics in resolving the
problems of CDM on small scales); in addition, the
possibility of more complex dark matter physics 
(decays, self-interactions, warm/mixed dark matter, e.g. \cite{nsdm0,nsdm1,nsdm2,nsdm3,nsdm4}) further complicate the comparison between
theory predictions and observations.  

Additional complications in comparing the results of $\Lambda$CDM
simulations to observations arise due to a certain vagueness on where
one should place the boundary of a bound, virialized dark matter halo
in the CDM model.  Simulated CDM halos are fitted well (at least on
scales of interest for defining their outer radius) by the
Navarro-Frenk-White density profile \cite{nfwnobreak}, which however
does not have a clear outer boundary. Rather, it follows an $r^{-3}$
power-law profile at large radii, with the mass included within $r$
diverging logarithmically as $r \rightarrow \infty$. The usual
solution is to take the virial radius to be located where the average
density within that radius is a predetermined factor over the background matter density at the redshift of virialization. This idea comes from the spherical collapse model: if one starts with a spherical perturbation in an otherwise homogeneous $\Lambda$CDM universe and allows it to evolve as a closed universe according to its own Friedmann equation, then one can predict the time it would formally collapse to a singularity, and hence the background matter density at that time. If one additionally assumes that, at this moment of final collapse, the radius of the perturbation, instead of becoming zero, is stabilized at the value predicted by the virial theorem, then one can also calculate the mean density of the perturbation at that time. The ratio of the two densities has a well-defined value, always equal to $18\pi ^2$ in an $\Omega_m = 1$ universe, or, in an $\Omega_m+\Omega_\Lambda=1$ universe, slowly increasing from $18\pi ^2$ at early times (before vacuum energy significantly affects the evolution of the expansion rate) to a value of $\sim 360$ at the present cosmic epoch \cite{all}. 

This approach involves two important ad hoc assumptions: that the time
of virialization of a structure is identical to the time of its formal
collapse to a singularity; and that the virial radius as calculated by
the recipe above is indeed a meaningful scale for a realistic bound
structure in the $\Lambda$CDM universe. Neither is obviously true. On
the one hand, virialization is a complex process that involves
interactions between dark matter particles, as well as between dark
matter substructures.  The collapse-to-a-singularity timescale is the
only calculable timescale in the spherical collapse model so it is the
first-recourse choice, but virialization itself cannot be described by
the spherical collapse model, and it does not have to operate on its
timescale. At best, the time of true virialization can only be
associated with the collapse-to-singularity time within a factor of a
few. On the other hand, cosmological simulations show that simulated
dark matter halos appear themselves unaware of their own virial
radius: their radial density profile slope is unchanged below and above the virial
radius as calculated above, and no break appears at radii a factor of
a few larger or smaller than the virial radius either
\cite{nfwnobreak, tavio}. In addition, the ``virial'' radius predicted by the
spherical-collapse recipe does not correspond to the spatial extent of
relaxed particles \cite{cuesta}.  Observational constraints on the
extent of bound matter on galactic scales similarly show discrepancies
with the usual definitions of the virial radius \cite{shull}.

 Given the considerations above, the question then arises: {\em are} there any predictions of $\Lambda$CDM regarding cosmic structure that are “clean” of baryonic physics and robust against the details of the way the predicted quantities are defined or calculated? 

In this work, we argue that such a prediction does exist, and it
involves the  {\em turnaround radius} of a bound structure - the
radius separating infalling or virialized material from the outwards
expansion of the Hubble flow. In contrast to the virial radius, this
is a well-defined structure boundary in simulations,
analytic calculations, and the observable Universe: if we calculate
the spherically averaged radial velocity of matter $\langle v_r \rangle$ around a center of gravity,
starting from spheres of very large radius and proceeding towards
smaller scales, the turnaround radius corresponds to the first time 
where $\langle v_r \rangle$ becomes zero.

Here we show that the well-established result of $\Lambda$CDM
structure formation that growth of structure does not continue
indefinitely (e.g., \cite{nl03, all,all2,all3,all4,all5,barrow,pfd,pfd2}) can be cast as a robust {\em testable} prediction: that the turnaround radius can never exceed an absolute upper bound, which, in an $\Omega_m+\Omega_\Lambda=1$ universe, is independent of time, and only depends on its mass and the value of the cosmological constant. This bound corresponds to the distance from the center of a bound structure  where the influence of matter and dark energy are at a balance.  If the mass shell at this radius has not reached turnaround yet, it never will, and neither will any mass shells at larger radii. The universe outside that bound will eventually accelerate away, while the universe as observed by inhabitants of the structure will asymptotically approach the Einstein static universe.

The bound can be tested observationally by estimating independently
the turnaround radius and enclosed mass of various structures. The
predictions of many alternative to $\Lambda$CDM cosmological models
for the same bound (if one exists at all) are diverse enough that the
bound can be used as an important cosmology discriminator.  Such
observations do exist for a few objects, demonstrating the feasibility
of such an endeavor. However, because the bound has not been recognized as a testable cosmological prediction up to now, these observations have not been systematically pursued. {\em We therefore urge for the undertaking of such observational efforts.}

\section{Qualitative considerations: why does the turnaround radius
  have a bound?}\label{qualitative}

Structure formation in $\Lambda$CDM cosmology is a cosmic battle
between expansion momentum, gravitational attraction, and dark energy. On small enough
scales, isolated, bound, relaxed gravitating systems are no longer adjusting
their radius\footnote{Before the effect of $\Lambda$ quenches
  structure growth, the mass of non-isolated cosmic structures changes through accretion and minor and major mergers; see e.g. \cite{belli} for a recent
discussion}; on very large scales, space and matter follow the
average expansion of the Universe. The scales in between, however, warrant more
careful consideration \cite{np71}.  

Collapsed structures grow by accreting surrounding mass shells: the
gravitational attraction of the central object will slow down a larger, co-centric mass
shell and detach it from the Hubble flow, causing it to turn around and
eventually collapse onto the central object. In a $\Lambda$CDM
universe, however, this process can only operate as long as the effect
of dark energy inside the mass shell does not overcome that of
matter. The dark energy associated with a cosmological constant does
not change with the scale factor $a_p$ of the perturbation, while
matter density falls as $a_p^{-3}$: as the scale factor of a mass
shell grows, the effect of dark energy within it grows relative to that of
matter. If the shell fails to reach turnaround by the time the two
effects become equal, dark energy will win, and the shell will start
expanding at an accelerated rate. 
The mass $M$ enclosed by the last shell that manages to reach
turnaround will be the {\em final mass} of that particular bound
structure. The radius at turnaround of this last shell is the maximum
turnaround radius, $R_{\rm ta,max}$, of mass $M$. 
The universe outside $R_{\rm ta,max}$ will eventually accelerate away,
while the universe as observed by inhabitants of the structure inside
$R_{\rm ta,max}$  will asymptotically approach the Einstein static
universe. The last shell to turnaround will grow increasingly slowly
and will only reach its final size, $R_{\rm ta, max}$, after an
infinite time. 

In the following section we will derive the relation between $R_{\rm ta,max}$ and $M$. 

\section{Quantitative derivations of the turnaround radius bound}

The value of the bound  is  robust against the detailed assumptions of its calculations, as we demonstrate by following through three different lines of derivation.
First, we consider the motion of a test particle under the influence
of a point mass in de Sitter space (in which case the mass and
$\Lambda$ are the only ingredients of the universe) and calculate the
conditions under which, when left from rest, the particle will fall
toward the mass, or it will be dragged away by the influence of
$\Lambda$. Next, we follow the evolution of a spherical density
perturbation within an otherwise homogeneous and isotropic universe of
$\Omega_m+\Omega_\Lambda=1$, from very early times when it co-expands
with the background universe, until the moment of turnaround (the usual
setup for the spherical collapse model, where homogeneity is assumed
both for the evolving perturbation as well as for the background
universe). Finally, we use the time derivative of the Friedmann
equation to seek a point of (unstable) equilibrium, with
not only zero expansion rate but also zero acceleration. 

\subsection{Spherical mass in de Sitter space}
The spacetime outside a spherically symmetric matter distribution with mass $M$ in a spatially flat de Sitter 
background with cosmological constant $\Lambda>0$ is described by 
the Schwarzschild - de Sitter (SdS) metric \cite{vittie,vittie2,vittie3}, which in static coordinates is 
$
d\tau^2=A(r) dt^2 - dr^2/A(r) -r^2 d\Omega^2\,,
$
with 
$A(r)=1-2M/r-\Lambda\, r^2/3$
in units with $G=1=c$, where 
$M$ is the mass
of the central body, 
$t$ the cosmic time and $r$ the physical distance from the center of
mass. 

The trajectory ($t(\tau), r(\tau), \theta(\tau)$, $\phi(\tau)$) of a test particle parametrized by the proper time $\tau$, 
satisfies 
\be
A(r)\, {\dot t}^2 - \frac{{\dot r}^2}{A(r)} - r^2\, {\dot\theta}^2 - r^2 \sin^2\theta\, {\dot\phi}^2 = 1
\label{L=1}
\ee
\be
A(r)\, \dot t = \eta \,,\; \frac{d}{d\tau}(r^2 \dot\theta)=r^2 \sin\theta \cos\theta\, {\dot\phi}^2 \,, \;
r^2\sin^2\theta\, \dot\phi=\lambda 
\label{rest}
\ee
where $\eta$ and $\lambda$ are 
fixed by the initial conditions and the ``dot" 
denotes differentiation with respect to $\tau$.
For radial motions 
($\lambda=0$)
and initial conditions $r(\tau_0)=r_0$ and
${\dot r}(\tau_0)=u_0$ we obtain 
\be
\frac{1}{2} {\dot r}^2 + U_0(r) = {\mathcal E}_0 
\label{radialeqn2}
\ee
where
$
U_0(r)=-M/r-\Lambda r^2/6 \,\; {\rm and}\;\, {\mathcal E}_0=(\eta^2-1)/2 = u_0^2/2 + U_0(r_0)
$.

The force acting on a unit test mass is given by 
$
F=-dU_0/dr =- M/r^2 + \Lambda r/3.
$
It vanishes at the critical radius $r_c =(3M/\Lambda)^{1/3}$
or, in physical units for
\be
r_c \equiv \left(\frac{3\,GM}{\Lambda \,c^2}\right)^{1/3}\,,
\label{rc}
\ee 
while $F$ is repulsive (attractive) for $r>r_c$ ($r<r_c$).
Thus, a test particle left from rest at $r_0>r_c$ will be accelerated away from the spherical mass $M$, 
dragged by the accelerated expansion of the universe. For $r_0<r_c$ it will fall on the mass $M$, while 
$r_0=r_c$ is an unstable equilibrium position.

\subsection{Spherical collapse of early-Universe perturbation}
The treatment of a single spherical mass inside otherwise empty de
Sitter space raises the question: is the limit expressed by
Eq.~(\ref{rc}) only applicable at very late times in the evolution of a $\Lambda$CDM
universe, where the accelerated expansion leads to a matter
distribution approximating the idealized configuration of the previous
section? In this section we show that this is not the case. If instead
we consider the evolution of a spherical overdensity from the early
universe (where its density contrast with the background universe is
extremely low) until the time it reaches turnaround, we derive exactly
the same limit. 

The evolution of a spherical perturbation of any size and sign in $\Lambda$CDM cosmology is governed by the ratio of the Friedmann equations for the perturbation and the background universe \cite{pfd,pfd2}, 
\begin{equation}\label{genl}
\left(\frac{da_{\rm p}}{da}\right)^2 = 
\frac{a_{\rm p}^{-1}+\omega a_{\rm p}^2-\kappa}{a^{-1} + \omega a^2}
= \frac{a}{a_{\rm p}}\frac{\omega a_{\rm p}^3-\kappa a_{\rm
    p}+1}{\omega a^3 +1}
\,,
\end{equation}
where $a$ is the scale factor of the Universe, $a_p$ is the scale factor of the perturbation, $\kappa$ characterizes the amplitude of the perturbation, and $\omega = \Omega_\Lambda/\Omega_m$ for the background Universe. The smallest $\kappa$ for which $\omega a_{\rm p}^3 -\kappa a_{\rm p} +1=0$ has a real positive solution (i.e. $a_p$ turns around) is 
$\kappa_{\rm min, coll} = 3\omega^{1/3}/2^{2/3}$.  The corresponding solution is the maximum turnaround radius, 
$a_{\rm p,ta,max}=(2\omega)^{-1/3} = \left(4\pi G \rho_m/\Lambda c^2\right)^{1/3}.$ 
All other collapsing
overdensities will have $a_{\rm p,ta}<a_{\rm p,ta,max}$.
The physical radius of a perturbation is  $R_{\rm p} =a_{\rm p} \left(3M/4\pi \rho_m\right)^{1/3}$, where $\rho_m$  is the matter density of the background Universe. 
We thus obtain
\begin{equation}
R_{\rm p,ta,max}=\left(\frac{4\pi G \rho_m}{\Lambda c^2}\right)^{1/3}\left(\frac{3M}{4\pi \rho_m}\right)^{1/3} =  
\left(\frac{3GM}{\Lambda c^2}\right)^{1/3}
\,,
\end{equation}
which is independent of cosmic time. 

\subsection{Zero Acceleration Mass Shell}

Having shown quantitatively that the limit exists and is the
same in both idealized configurations we considered in the previous
sections, we can express it in a fashion that unifies conceptually the
two approaches. 

Ultimately, the upper bound to the turnaround radius separates two qualitatively different regions in spacetime. For smaller radii, any mass shell at rest will accelerate inwards. For larger radii, even a mass shell at rest will accelerate outwards, due to the dominating influence of vacuum energy. The maximum turnaround radius mass shell, therefore, must be a zero-acceleration one: it should define a sub-universe obeying its own Friedmann equation, the time derivative of which satisfies $\ddot{a_p}=0$ (where $\ddot{a_p}$ is the second time derivative of the scale factor of the perturbation) when the matter density inside it is equal to $3M/4\pi R_{\rm ta,max}^3$. 

The time derivative of the Friedmann equation can be written as
\begin{equation}
\ddot{a_p} = -4\pi Ga_p(\rho c^2 +3p)/3. 
\end{equation}
Therefore, the no-acceleration condition then yields 
\begin{equation}\label{condition}
\rho c^2 +3p = 0
\end{equation}
where in our case $\rho = \rho_v +\rho_m$. For a matter term heavily dominated by pressureless dark matter, the only source of pressure is vacuum energy. But vacuum energy (assuming it behaves as a cosmological constant) satisfies $\rho_v c^2 =-p_v$. Equation (\ref{condition}) then becomes 
\begin{equation}\label{densities}
\rho_m  = 2\rho_v
\end{equation}
 or  $3M/4\pi R_{\rm ta,max}^3 = 2\Lambda c^2/8\pi G$
which yields the familiar by now result
\begin{equation}
R_{\rm ta, max} = \left(\frac{3GM}{\Lambda c^2}\right)^{1/3}\,,
\end{equation}
which is independent of cosmic time. 

As it is obvious from Eq.~(\ref{densities}),  an alternative way to state the limit is in terms of the mean matter
density within the turnaround radius, $\rho_{\rm m,ta}$, which, from
Eq.~(\ref{densities}), has to satisfy $\rho_{\rm m,ta}\ge 2\rho_v$. If
the vacuum density within a spherical region is higher than one-half
the matter density within that region at any given time, then the mass
shell at the edge of that region {\em has} to be expanding, and it
will continue to do so indefinitely at an accelerated rate. 

For a cosmology satisfying $\Omega_m(a)+\Omega_\Lambda(a)=1$, (where
$\Omega$ denotes a density in units of the critical density $\rho_{\rm
crit}(a)$ at a
specific cosmic epoch labeled by the cosmic scale factor $a$),
we can express this limit in terms of the density contrast of a
non-expanding spherical region with respect
to the background matter density $\rho_{m,b}$, $\delta_{\rm ta}=\left(\rho_{m,
   \rm  ta} - \rho_{m,b}\right)/\rho_{m,b}$:
\begin{equation}\label{delta}
\delta_{\rm ta} = \frac{\rho_{\rm m, ta} }{\rho_{m,b}} -1 
\ge \frac{2\rho_v}{\rho_{m,b}} - 1= \frac{2\Omega_\Lambda(a)}{\Omega_m(a)} -1 =\frac{2}{\Omega_m(a)} - 3\,.
\end{equation}
For the present cosmic epoch, the best estimate for $\Omega_m$ is $0.315$ \cite{planck} so inequality
(\ref{delta}) yields $\delta_{\rm} \ge 3.35$: although the limit is applicable to
non-expanding structures, it is relevant for densities that are 
more than 100 times smaller than the virial overdensity at the
present cosmic time \cite{all}.

 It is also clear that
if $\Omega_{m} (a) >2/3$ then inequality (\ref{delta}) is satisfied for any
overdensity (positive density contrast), and therefore the bound can
only be used as a cosmological discriminator at late times in the
evolution of the Universe. The redshift at which $\Omega_m(a)=2/3$ can
be found from 
\begin{equation}\label{omegas}
\Omega_m(a) = \frac{\rho_{m, b, \rm present} a^{-3}}{\rho_{\rm crit,
    present} \left[\Omega_\Lambda + \Omega_m a^{-3}\right]} =
\frac{\Omega_m a^{-3}}{\Omega_\Lambda + \Omega_m a^{-3}}\,.
\end{equation}
For cosmological parameters from
\cite{planck}, Eq.~(\ref{omegas}) gives $a=(1+z)^{-1} =0.613$, 
and therefore meaningful observational tests of the bound must focus
at redshifts $<0.63$. That at early times the bound does not come into 
play is qualitatively expected: at high redshifts where $\Lambda$ is
very subdominant, structure formation should proceed in a manner
indistinguishable from the $\Omega_m=1$ cosmology, which predicts no bound.

\section{Testing the bound observationally} The maximum turnaround
radius is a robust, time-independent prediction of $\Lambda$CDM
cosmology. The only assumption that enters its derivation is that of
spherical symmetry. This prediction states
that, in a $\Lambda$CDM universe, there can never be a {\em
  nonexpanding} structure whose radius is greater than
$\left(3GM/\Lambda c^2\right)^{1/3}$, where $M$ is the structure's
mass. In this section, we discuss how this prediction can be used as
an observational test of various cosmological models.  

\subsection{What observations are necessary?}\label{whatobs}

Observational tests of the bound would aim to answer the question:
is the turnaround radius of a given structure less than
$\left(3GM/\Lambda c^2\right)^{1/3}$? This requires {\em independent}
measurements of the turnaround radius around a center of mass, and the
total mass enclosed within said radius. 

The turnaround radius around a center of mass is a well-defined,
although not always easy to measure: it is the radius where the
average radial velocity with respect to said center is zero, while it is positive for all radii
larger than the turnaround radius. The last part of the definition is
necessary to distinguish between the accretion boundary and the
turnaround radius in structures that experience infall: the average radial velocity changes from zero to
negative as the accretion boundary is crossed outwards, and from zero
to positive as the turnaround radius is crossed outwards. 

The turnaround radius can be determined by high-precision observations
of the Hubble flow in cosmic structures that consist of at least
several galaxies. Such observations have been successfully pursued for
a number of local-universe structures (typically galaxy groups or
nearby clusters, e.g. \cite{TALG, kk, TAcenA, TAvirgo, TAeridanus, TAvirgonew}). The uncertainty with
which the turnaround radius (commonly referred to as  the
``zero-velocity surface'') has been determined in these studies is
between $10-30\%$. 

The mass within the turnaround radius is more difficult to
measure. Estimates of the turnaround radius are typically used in
combination with predictions of the spherical collapse model in the
preferred cosmology at the time of each study to obtain an estimate of
the structure mass. This approach clearly enforces by construction the
validity of the bound we propose to test here, so a different way to
estimate the mass is needed. Summing up the masses of the
constituent galaxies \cite{li}, lensing studies, or X-ray spectroscopy studies
\cite{Xlens} could
provide such alternatives. 

A problem with such independent methods of mass estimation is that
they usually do not correspond to regions extending out to the
turnaround radius. This is partly a matter of observational strategy
(the possibility of using mass measurements out to the turnaround
radius as a test of cosmology has not been recognized up to now);
often however it is a matter of sensitivity: the X-ray brightness of
galaxy clusters, for example, would in most cases be too low to
detect at the turnaround radius. 

The question then arises: would it be possible to use such estimates
of mass - independent from the determination of the turnaround radius,
but only ``counting'' mass out to a smaller radial boundary - as tests
of the bound? The answer is yes. Such a mass estimate constitutes a
lower limit (let us call it $M_{\rm ll}$) to the total structure mass (let
us call it $M_{\rm tot}$) out to the turnaround
radius. If the measured turnaround radius of the structure does not exceed the
bound corresponding to a mass $M_{\rm ll}$, then it is not going to
exceed the bound corresponding to its true total mass $M_{\rm tot}$
either, since the value of bound increases monotonically with
mass. If, on the other hand, the bound corresponding to $M_{\rm ll}$
{\em is} exceeded, it is not guaranteed that the bound corresponding
to $M_{\rm tot}$ will also be exceeded. Such a structure can be
labeled as a candidate for bound violation, and targeted, more sensitive follow-up
observations can be undertaken to determine the mass $M_{\rm tot}$
(preferably by more than one methods), to either confirm or reject a
bound violation together with its cosmological implications. 

Similarly, some information can be extracted from structures for which
good mass estimates of their mass out to some radius exist, but for
which a reliable estimate of their turnaround radius is not
possible. The bound discussed here is an absolute upper limit
for all non-expanding shells around a center of gravity in
$\Lambda$CDM cosmology. For this reason, {\em any} mass shell not
moving away from the structure center cannot have a radius larger
than $\left(3GM/\Lambda c^2\right)^{1/3}$. An upwards violation of this limit
by a non-expanding shell internal to the turnaround radius is still a
violation of $\Lambda$CDM cosmology. In contrast, a mass/radius
measurement that obeys the bound for a shell internal to the
turnaround radius is non-informative. 

\subsection{How many structures need to be observed?}

A single high-confidence observation of the turnaround radius of a
structure of known mass exceeding the bound would be sufficient to
constrain $\Lambda$CDM cosmology. However, even if the cosmological
model differs from $\Lambda$CDM in such a way that the bound
does not hold, not every structure will have a turnaround radius
larger than $\left(3GM/\Lambda c^2\right)^{1/3}$. The reason is the
statistical nature of structure formation: different regions
(enclosing similar mass) start off with slightly
different average overdensities in the early universe and will have
reached a different stage in their evolution by a certain redshift. As
a result, the more structures we observe (estimate their turnaround
radius and enclosed mass), the higher the probability to identify structures
exceeding the bound (if such a violation is allowed by the actual
cosmology of our Universe).

The number of structures that need to be observed in order to achieve
a certain probability of encountering a bound-exceeding object depends
on the cosmological model - alternative to $\Lambda$CDM - that would be
producing such violations. For example, if the perceived acceleration
of the expansion of the Universe were a result of a local large-scale
inhomogeneity rather than a cosmological constant, as suggested, e.g.,
by \cite{nobound1,nobound2,nobound3,nobound4}, then there should be no bound
to the turnaround radius - even regions with very low
overdensities in the early universe are allowed to turnaround and
eventually collapse. Because in the standard $\Lambda$CDM model the
acceleration of the Universe has largely suppressed structure
formation by the present cosmic epoch (and thus almost all structures should
be close to their ultimate size by now,  e.g. \cite{nl03, all2}), the
observation of only a few ($\lesssim 10$) mass/turnaround radius pairs
should be adequate to identify bound-violating objects. 

For other cosmological models that also predict bounds but of a
different amplitude, the number of sources that need to be
observed for a certain probability of exceeding the bound can be
calculated on a case-by-case basis. In general, the looser the bound
of the ``true'' cosmology, the higher the chance to identify a
bound-violating structure with a fixed number of observations. 

A specific example of a cosmological model resulting in a bound of different
amplitude is that of dark energy having a general equation of state
parameterized by 
$w=p/\rho<-1/3$. In this case, the maximum turnaround radius is \cite{ptt}
\begin{equation}
R_{\rm ta, max} = \left( -\frac{3M}{4(1+3w)\pi \rho_E}\right)^{1/3}\,,
\end{equation}
which gives the same result as Eq.~(\ref{rc}) for $w=-1$. The bound is
tighter the more negative $w$ is. As discussed in \cite{ptt},
existing data may already indicate violations of the bound for
$w<-1$. Several
structures potentially violating the bound for $w=-2.5$ (which is
tighter than the ``true'' bound by about a factor of 2, assuming that
$\Lambda$CDM is the ``true'' cosmology) are
found among observations of fewer than 10 objects. 

\subsection{Which structures should we be observing?}

Because the bound is independent of cosmic time, while structures
continue to grow with time and approach their maximum possible size,
the best cosmic epoch to test the bound observationally is the present
one. 

In a standard hierarchical structure formation scenario, smaller
structures form first (have a higher chance to begin with high
overdensities in the early Universe), and are also the first to reach
their maximum sizes. For this reason, the optimal mass range to target
in order to possibly identify bound-violating structures would be the
smallest masses for which the turnaround radius can be confidently
determined. Assuming that the turnaround radius estimate would be done
using the Hubble-flow analysis employed in past studies
(e.g. \cite{kk}), then we need several galaxy-members within the
target structure, so the natural class of targets would be groups of
galaxies. 

More accurate estimates of the 
range of masses that hold the maximum discriminating power (i.e. that
are likely to have reached their maximum attainable mass at the
present cosmic epoch while at the same time being large enough to
allow confident determination of their turnaround radius) for various
cosmological models can be
produced through simulations and semi-analytical models.  

\subsection{Can such observations produce ``false positives''?}

If, then, $R$ or, equivalently,  $\rho$ that violate the limit {\em
  are} measured, the discrepancy cannot be attributed to baryonic
physics (which is only a very small perturbation at turnaround
scales), complexities of the process of collapse and dynamical
relaxation, or uncertainties in the way radii are defined. The only
simplifying assumption entering its calculation is spherical symmetry,
and accounting for deviations from sphericity is not expected to
change this result significantly, as a mass shell  at its moment of
turnaround has not undergone any contraction yet. However, if the
bound is to be used to confidently constrain the cosmological model,
the magnitude of any non-sphericity effects has to be determined
accurately. 

The effect of non-sphericities on turnaround and the ultimate radii of
cosmic structures has been discussed analytically by \cite{barrow} and, to
some extent, in cosmological simulations by \cite{all2}. In the
latter work, it was found that, for realistic structures with
mass between $10^{14}$ and $10^{16} {\rm M_\odot}$, the “sphere of
influence” is no more than 30\% larger than the
what would be expected in spherical symmetry. Ideally, the problem should be addressed by 
analysis of cosmological simulations targeting the mass ranges of
interest and using the same way to define the turnaround radius in simulations and observations of asymmetric
structures.

\subsection{How accurate need the observations be?}

Assuming in the future an analysis tailored to the purposes of the
test proposed here provides independent measurements for both
mass and turnaround radius of a sample of structures large enough so
as to include at least one structure with mass very close to its {\em
  final mass} (see \S \ref{qualitative}), we give below a very simple
estimate of the
confidence at which various levels of violation of the bound can be
established, given a certain accuracy for measurement of structure
mass and radius. 

Let us assume that a structure of final mass $M$ exceeds the bound by an amount
$\Delta R$, i.e. its radius $R$ is equal to 
\begin{equation}
R=R_{\rm ta, max} +\Delta R = \left( \frac{3GM}{\Lambda
    c^2}\right)^{1/3} +\Delta R.
\end{equation}
The question that we would like to answer is the following: for a
given fractional violation of the bound, i.e. for a given value of
$\lambda = \Delta R/R_{\rm ta, max}$, what is the uncertainty in the
measurements of $R$ and $M$ we can tolerate, to establish the
violation of the bound at a given confidence level (number of $\sigma$)? 
We will denote the number of $\sigma$ with $n$, the uncertainty in the
measurement of radii with $\sigma_R$, and the uncertainty in the
measurement of masses with $\sigma_m$. 
We measure $\Delta R$ by measuring $R$ directly, and by estimating
$R_{\rm ta,max}$ indirectly, using a mass measurement: $\Delta R = R - R_{\rm ta, max}$. Simple error
propagation implies that 
$\sigma_{\Delta R}^2 = \sigma_R^2 + \sigma_{R_{\rm ta,max}}^2\,.$
Similarly, $\sigma_{R_{\rm ta,max}}^2$ depends on the
uncertainty in our mass measurement 
$\sigma_{R_{\rm ta,max}}/R_{\rm ta,max}= (1/3)\sigma_m/M,$
where we have ignored the -much smaller- error entering through
$\Lambda$. Combining these two results, we obtain 
\begin{equation}\label{uncertainty1}
\sigma_{\Delta R}^2 = \sigma_R^2 +\frac{1}{9}\sigma_m^2\frac{R_{\rm ta, max}^2}{M^2}\,.
\end{equation}
Using 
\begin{equation}
n=\frac{\Delta R}{\sigma_{\Delta R}} = \lambda \frac{R_{\rm ta,max}}{\sigma_{\Delta R}}
\end{equation}
we can write Eq.~(\ref{uncertainty1}) as 
\begin{equation}\label{uncertainty2}
\lambda^2 = n^2\left(\frac{\sigma_R^2}{R_{\rm
      ta,max}^2}+\frac{1}{9}\frac{\sigma_m^2}{M^2}\right). 
\end{equation}
Equation (\ref{uncertainty2}) implies that  a few percent accuracy in the
measurements of $R$ and $M$ allows us to establish a $10\%$ bound
violation at the 3$\sigma$ level, and a $6\%$ violation at 
$2\sigma$. If we were to simply increase our sample of structures with
independently measured $R_{\rm ta}$ and $M$ {\em without any
  improvement in measurement techniques} (i.e. with uncertainties similar
to the ones quoted for the few cases discussed above, $\sigma_m/M \sim 30\%$,
$\sigma_R/R \sim 5\%$), for a structure close to its final mass
we can establish a $30\%$ bound violation at $3\sigma$, and we can
already detect a $1\sigma$ tension between data and prediction for a
$10\%$ violation. An improvement in mass measurements would at this
point have the largest impact in terms of updating observational
techniques. 

Regarding the effect of non-sphericities, if it can be simply represented
as a multiplicative factor $f$ on the spherically symmetric result,
i.e.
\begin{equation}
R_{\rm ta, max, actual} = fR_{\rm ta,max, \, spherical \, \, symmetry}
\end{equation}
then the results above still hold, as long as $\lambda$ describes a fractional
violation over the actual bound, including any non-sphericity
corrections. As a result, an increase in accuracy with which deviations from
spherical symmetry are accounted for can be best achieved with systematic analysis of ultimate
turnaround radii as a function of mass in simulations, with the radii defined as
closely as possible to the way they can be measured in observations;
the observational accuracies required are still those quoted
above. 

\subsection{What can already existing data tell us?}

From the latest determination of the cosmological parameters by
Planck \cite{planck}, the present estimate of
the matter density parameter is $\Omega_m = 0.315\pm0.017$ so
$\Omega_\Lambda = 1-\Omega_m=0.685$. For this value of
$\Omega_\Lambda$, the bound becomes  $R_{\rm ta, max} = 11.2\pm 0.1
{\rm \, Mpc} \left(M/10^{15} M_\odot\right)^{1/3}$.  In the discussion
below, whenever the value of the Hubble
parameter was needed to obtain a mass or radius in physical units for
specific objects, we have used  $H_0 = 67.3 {\rm \,
  km \, s^{-1} Mpc^{-1}}$ \cite{planck}. 

A few sanity checks yield comforting results. For the solar system, a central mass
of 1 ${\rm M_\odot}$ yields a bound of $\sim$ 100 pc, much greater
than the solar system radius. The sun would be able to hold onto his
own even if the hand of some angry deity were to cast the solar system
in the middle of a cosmic void. 
For the Milky Way, assuming that the Leo I dwarf spheroidal at a distance of $254 \pm 16 $kpc is bound to the Galaxy \cite{leo,leo2},  $M= 1-2.4 \times 10^{12} {\rm \, M_{\odot}}$ yielding a  bound of $1.1 - 1.5 {\rm \, Mpc}$, much larger that the Leo I distance. 

\begin{figure}
\begin{center}
\includegraphics[width=.8\textwidth, clip=true]{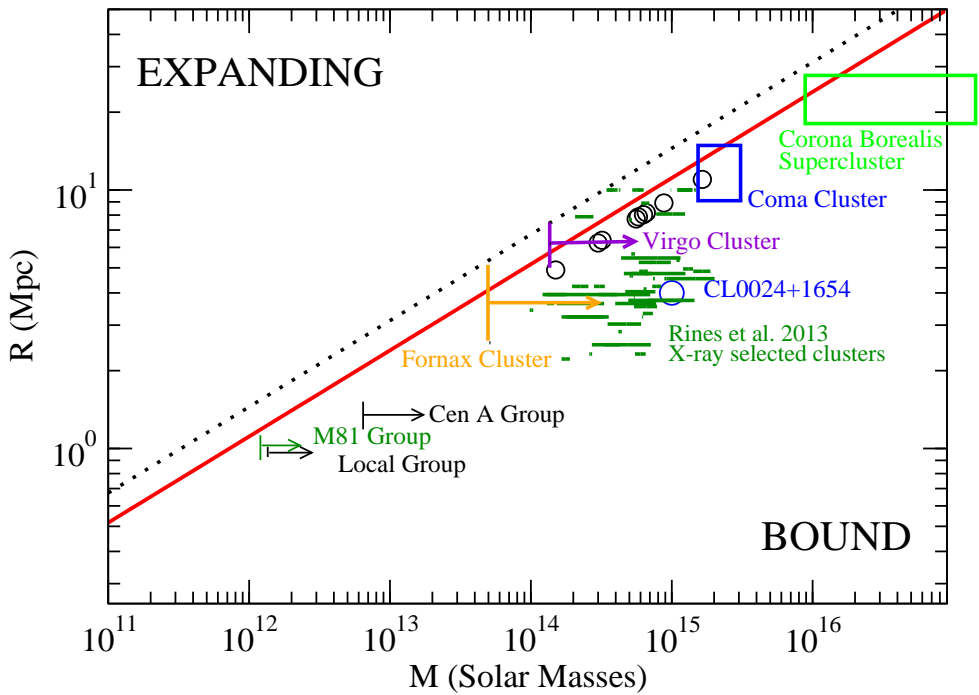}
\end{center}
\caption{\label{thefig} Data masses and turnaround radii for several
  local-Universe structures compared to the $\Lambda$CDM turnaround
  radius bound (red line). The dotted line represents the location of
  the bound accounting for the estimated effect of non-sphericities. See text for details. }  
\end{figure}

In Fig. 1 we have compiled data from the literature on the radii and
masses of groups of galaxies, clusters of galaxies, and galaxy
superclusters, and we plot them together with the turnaround radius
bound: with the red solid line, we plot the spherically 
symmetric result, derived in this work; with the black dotted line we
plot our estimate (using the simulation results of \cite{all2}) of the
bound including the effect of deviations from 
spherical symmetry. 

Because the analyses that resulted in this data were
not tailored to the test we wish to perform here, in most cases we do
not have independent estimates of the turnaround radius and the mass
enclosed out to the turnaround radius. For example, \cite{rines03}
calculate mass profiles for clusters A119, A168, A496, A539, A576, 
A1367, A1656, A2199 and A194 using the caustic technique, and
then use a constant-overdensity criterion to calculate the location of
the turnaround radius. Although such data cannot be used to test the
bound (since the radius and mass estimates are derived from each
other), we plot this data with open black circles for illustration
purposes in Fig. 1. As expected, they lie on a line parallel to that
defined by the bound, as both the analysis of \cite{rines03} and
the bound correspond to a constant matter density. In contrast, we
also plot in dark green a
sample of 58 X-ray selected clusters from \cite{rines2013}  that also had their mass measured
using the caustic technique, out to a maximum radius where caustics
were detected. This data populates almost an order of magnitude in
radius for similar masses, as expected for a set of structures at
different evolutionary stages (as clusters of galaxies are believed to
be in the current cosmic epoch). Five of these structures (A1361, A1033,
A1235, A646, A1366) exceed the
spherically symmetric bound but not the estimated bound when
non-sphericities are included; and two structures (A2055, A1918) reach the latter as
well.

In other cases, the turnaround radius is directly obtained by
examining the Hubble flow and the peculiar velocities around a given
cosmic structure (e.g., 
\cite{kk, TALG, TAcenA, TAvirgo, TAeridanus, TAvirgonew}). These
estimates are typically used to derive the enclosed mass based on
spherical collapse model fitting. Clearly, this approach enforces by
construction the validity of the bound we propose here to test, so
these mass estimates cannot be used in combination with the turnaround
radius measurements to test the bound. What we do instead is look for
independent estimates of the mass of the structures in question. These
generally do not estimate the mass out to the turnaround radius, but
rather out to some smaller radial scale. These mass estimates then
represent {\em lower limits} on the mass enclosed within the
turnaround radius, and have been plotted as such in Fig. 1. We have
included such measurements for:
\begin{itemize} 
\item  the Local Group (turnaround radius from
\cite{TALG}; mass estimate: lower limit on the sum of the masses of
M31 and the Milky Way, as estimated by \cite{andrey}); 
\item the M82 group of galaxies (turnaround radius from \cite{ketal};
  mass estimate: sum of virial masses of members \cite{ketal}). 
\item the Cen A group of galaxies (turnaround radius from
  \cite{TAcenA}; mass estimate: member orbital mass \cite{TAcenA}); 
\item the Fornax cluster (turnaround radius from \cite{TAeridanus};
  mass estimate: caustic technique within a smaller radius of 1.4 Mpc
  \cite{fornaxmass});
\item  the Virgo cluster (turnaround radius from \cite{TAvirgonew};
  mass estimate: X-ray spectroscopic mass within smaller radius of 1.2
  Mpc \cite{virgo}).
\end{itemize}

In Fig. 1 we also plot, with the open blue circle, data for the mass
of CL0024+1654, obtained through lensing studies \cite{kneib}. In this
case, radius and mass match, but the radius is not the turnaround
radius. The point does not exceed the bound so, as discussed in
Section \ref{whatobs}, such an observation is non-informative. 

For the Coma cluster, we use data from the study of its infall
region by \cite{rines01}. The range of masses corresponds to the sum
of the masses calculated by profile fitting within and outside the
virial radius; the possible location of the turnaround radius is
estimated from the location of the caustics in redshift space. 

One of the two largest known structures in the local universe is the Corona Borealis
supercluster (the second being the Shapley supercluster which, however,
appears not to have reached turnaround at present \cite{shapley}, and the mass/size
of which is therefore not constrained by the turnaround radius
bound). Whether the Corona Borealis supercluster {\em as a whole} has reached
turnaround at present is still a matter of debate  \cite{borealis2};
however, the latest data and analysis of \cite{borealis} indicate that
this is likely the case, so we include data on its mass and radius in
Fig. 1. The mass range results from estimates by \cite{borealis} using
a variety of different methods (including the virial theorem, the
caustics technique, and orbital velocity/position fitting to the spherical
collapse model). The turnaround radius these authors derive is not
independent from their mass estimate (they use the spherical collapse
model to estimate a turnaround radius from their mass estimate);
instead, we use as lower and upper limits on the turnaround radius the
distances from the center of the supercluster of the farthest member
clusters that is collapsing towards the center, and the nearest member
cluster that is receding from it, respectively.

All of the structures in Fig. 1 are consistent within uncertainties with the maximum
turnaround radius limit. However, there are also a number of
structures that, within uncertainties, are also candidates for a
possible limit violation (although none exceed our estimate of the
bound when the effect of non-sphericities is included). 
Future observational or analysis efforts
seeking to test the limit at higher confidence could potentially focus
on these structures first.  

Different radii plotted in Fig. 1 have different degrees of proximity
to the exact definition of a turnaround radius as discussed in
Section \ref{whatobs}. The ``zero velocity surface'' of
e.g. \cite{TALG} is conceptually identical (``a surface which
separates the overdensity against the homogeneous cosmic expansion'',
practically the surface that separates galaxies with positive and
negative radial velocities compared to the group centroid). Even in
this case, however, differences might arise due ``real-world'' effects
(non-spherical structures, in which case, there is not enough
information to reconstruct the full velocity field from the accessible
line-of-sight velocity observations, and peculiar velocities) and the way this
surface is determined in practice (e.g. lack of a sufficient number of
objects close to the turnaround radius to determine the radius
accurately).  It is less clear how closely the ``maximum radius'' 
of \cite{rines2013} corresponds to the turnaround radius of a
structure. The most direct way to assess such results and improve the
comparison with theoretical expectations is by conducting mock
observations using numerical simulations. In such simulations, the exact turnaround
radius (as defined here) is known for every structure, and the values
that would be derived by observations for each observable radius (zero velocity
surface, maximum caustic radius etc) can also be estimated,
yielding dependable, mass-dependent estimates of any systematic
deviations. 

The structures that are plotted on Fig. 1
were selected based on data availability rather than proximity to the
limit; yet, all of them are close to the limit, indicating that
structure growth is reaching its end at the current cosmic
epoch. On the other hand,  we do not see any strong violations of the
limit among these $\sim 60$ structures (covering 4 orders of magnitude
in mass) either. These two facts together are an important,
{\em local} indication of dark energy at work.

All of the smaller structures (groups of galaxies) that we have
examined lie under the limit even within errors. If hierarchical
$\Lambda$CDM structure formation holds, this is most likely
the effect of more accurate observations: in this scenario, it is the smallest
structures that reach their limiting size first. Indeed, the
fractional errors in the determinations of the turnaround radius in
smaller structures are significantly smaller. If this is indeed the
case, as observations of larger structures improve, these points will
also move to the lower right of Fig. 1. This will be a strong argument
in favor of $\Lambda$CDM cosmology and against inhomogeneous no-$\Lambda$
 alternatives. 

On the other hand, if more
accurate observations of higher-mass structures confirm that these
structures are closer to the limit, or even violating it, this will constitute
evidence hard to reconcile with $\Lambda$CDM.

\section{Discussion}

That  essential to our
existence structures (such as the solar system or, by chemical-evolution arguments, the
Milky Way) do not lie close to the limit, prohibits the use of the
limit to provide an ``anthropic'' explanation to the low value of the
cosmological constant, consistent with the findings of
\cite{weinberg}. However, it has not escaped our notice that a strong
limit on the value of $\Lambda$ could be placed  by requiring that
measuring $\Lambda$ be feasible (see, e.g., \cite{coincidence}.)

Alternative cosmological models have different predictions for the
maximum turnaround radius, if a bound exists at all. For example, the
dark energy equation of state will affect the value of the maximum
(since no dark energy corresponds to no bound at all) \cite{ptt}. MOND
predictions would depend on the specifics of each theory, but they are
in practice guaranteed to be different from $\Lambda$CDM, while it is
not obvious that following the three lines of reasoning presented here
would even result to the same answer for a MOND-consistent gravity
theory (Skordis, private communication). Although a detailed treatment of the bound in
alternative cosmologies is beyond the scope of this letter, it is
clear that the bound can act as a cosmology discriminator.   

Even without improving the
accuracy of mass and radius measurements, a larger sample of
structures with independently observed turnaround radius and included
mass could give at least first indications for bound violations down
to $10\%$. We therefore urge towards such efforts,
which will challenge the limit and $\Lambda$CDM cosmology.

\acknowledgments

TNT would like to thank John Iliopoulos for useful discussions and his
suggestion to make these notes public and C. Skordis for insightful
comments on structure formation in the MOND framework. We thank Enea
Romano and Dimitrios Tanoglidis for stimulating discussions, and an
anonymous referee for constructive comments that improved this paper. 
This work was supported in part by European Union's Seventh Framework Programme under grant agreements 
(FP7-REGPOT-2012-2013-1) no 316165,
PIF-GA-2011-300984, PCIG10-GA-2011-304001, PIRSES-GA-2012-316788, the EU program ``Thales'' MIS 375734  and was also co-financed by the European Union (European Social Fund, ESF) and Greek national funds through the Operational Program ``Education and 
Lifelong Learning'' of the National Strategic Reference Framework (NSRF) under ``Funding of proposals that 
have received a positive evaluation in the 3rd and 4th Call of ERC
Grant Schemes'' and under the  ``ARISTEIA'' Action. 




\begin{thebibliography}{999}

\bibitem{larsb} Bergstr{\"o}m, L.\ 2012,  Annalen der Physik,
  524, 479
\bibitem{gianfranco}Bertone, G., Hooper, D., \& Silk, J.\ 2005, Phys.~Rep., 405, 279 

\bibitem{marusa} Brada{\v c}, M., Allen, S.~W., Treu, T., et al.\
  2008, \apj, 687, 959

\bibitem{marusa2}Clowe, D., Markevitch, M., Brada{\v c}, M., et al.\
  2012, \apj, 758, 128

\bibitem{markevich}Markevitch, M., Gonzalez, A.~H., Clowe, D., et al.\ 2004, \apj, 606, 819  

\bibitem{finkbeiner} Daylan, T., Finkbeiner, 
D.~P., Hooper, D., et al.\ 2014, arXiv:1402.6703 

\bibitem{bulbul} Bulbul, E., Markevitch, 
M., Foster, A., et al.\ 2014, arXiv:1402.2301 

\bibitem{andromeda} Boyarsky, A., 
Ruchayskiy, O., Iakubovskyi, D., \& Franse, J.\ 2014, arXiv:1402.4119 


\bibitem{famaey} Famaey, B., \& McGaugh, S., 2012, Living
  Rev. Relativity, 15, 10
\bibitem{milgrom}Milgrom, M., 2010, PoS HRMS2010, 033

\bibitem{angus} Angus, G.~W., Famaey, B., 
\& Diaferio, A.\ 2010, MNRAS, 402, 395 

\bibitem{planck} Planck 
Collaboration, Ade, P.~A.~R., Aghanim, N., et al.\ 2013, arXiv:1303.5076 

\bibitem{weinberg} Weinberg, S.\ 1987, Physical 
Review Letters, 59, 2607 

\bibitem{mondreview} Famaey, B., \& McGaugh, S.\ 2013, Journal of Physics Conference Series, 437, 012001 

\bibitem{baryons1} Brooks, A.~M., Kuhlen, 
M., Zolotov, A., \& Hooper, D.\ 2013, \apj, 765, 22 


\bibitem{baryons2} Arraki, K.~S., Klypin, 
A., More, S., \& Trujillo-Gomez, S.\ 2014, \mnras, 438, 1466

\bibitem{baryons3} 
Garrison-Kimmel, S., Rocha, M., Boylan-Kolchin, M., Bullock, J.~S., 
\& Lally, J.\ 2013, \mnras, 433, 3539 

\bibitem{nsdm0} Peter, A.~H.~G., \& Benson,
A.~J.\ 2010, \prd, 82, 123521

\bibitem{nsdm1} Vogelsberger, M., 
Zavala, J., \& Loeb, A.\ 2012, \mnras, 423, 3740 

\bibitem{nsdm2} Lovell, M.~R., Eke, V., 
Frenk, C.~S., et al.\ 2012, \mnras, 420, 2318 

\bibitem{nsdm3} Anderhalden, D., 
Schneider, A., Macci{\`o}, A.~V., Diemand, J., 
\& Bertone, G.\ 2013, \jcap, 3, 14 

\bibitem{nsdm4} Peter, A.~H.~G., Rocha, 
M., Bullock, J.~S., \& Kaplinghat, M.\ 2013, \mnras, 430, 105 


\bibitem{nfwnobreak} Navarro, J.~F., Frenk, C.~S., \& White, S.~D.~M.\
  1996, \apj, 462, 563 

\bibitem{all} Lahav, O., Lilje, P.~B., Primack, J.~R., \& Rees,
  M.~J.\ 1991, MNRAS, 251, 128

\bibitem{tavio}Tavio, H., Cuesta, A.~J., Prada, F., Klypin, A.~A., \& Sanchez-Conde, M.~A.\ 2008, arXiv:0807.3027 

\bibitem{cuesta} Cuesta, A.~J., Prada, F., Klypin, A., \& Moles, M.\ 2008, \mnras, 389, 385

\bibitem{shull} Shull, J.~M.\ 2014, \apj, 784, 
142

\bibitem[Nagamine 
\& Loeb(2003)]{nl03} Nagamine, K., \& Loeb, A.\ 2003, NewA, 8, 439 

\bibitem{all2} Busha, M.~T., et al.\ 2003, \apj, 596, 713

\bibitem{all3}Teerikorpi, P., Chernin, A.~D., Karachentsev, I.~D., \&
  Valtonen, M.~J.\ 2008, A\&A, 483, 383
\bibitem{all4} Eingorn, M., \& Zhuk, A.\ 2012, \jcap, 9, 26
\bibitem{all5}Eingorn, M., Kudinova, A., \& Zhuk, A.\ 2013, \jcap, 4, 10; 

\bibitem{barrow} Barrow, J.~D., \& Saich, P.\ 1993, \mnras, 262, 717

\bibitem{pfd} Eke, V.~R., Cole, S., 
\& Frenk, C.~S.\ 1996, \mnras, 282, 263

\bibitem{pfd2}Pavlidou, V., \& Fields, B.~D.\ 2005, \prd, 71, 043510; 

\bibitem{belli} Belli, S., Newman, A.~B., 
\& Ellis, R.~S.\ 2014, \apj, 783, 117 


\bibitem{np71} Noerdlinger, P.~D., \& Petrosian, V.\ 1971, \apj, 168, 1 

\bibitem{vittie} G.C. McVittie, Mon.Not.R.Astron.Soc. 933, 325 (1933)

\bibitem{vittie2}Chang Jun Gao, 2004, Class. Quant. Grav., 21, 4805,
 and earlier references therein

\bibitem{vittie3}Kaloper, N., Kleban, M., \& Martin, D.\ 2010, \prd, 81, 104044 



\bibitem{TALG}Karachentsev, I.~D., Sharina, M.~E., Makarov, D.~I., et
  al.\ 2002, \aap, 389, 812 

\bibitem{kk} Karachentsev, I.~D., \& Kashibadze, O.~G.\ 2006, Astrophysics, 49, 3

\bibitem{TAcenA} Karachentsev, 
I.~D., Tully, R.~B., Dolphin, A., et al.\ 2007, \aj, 133, 504 

\bibitem{TAvirgo} Karachentsev, I.~D., \& Nasonova, O.~G.\ 2010,
  \mnras, 405, 1075 

\bibitem{TAeridanus} Nasonova, O.~G., de Freitas Pacheco, J.~A., \& Karachentsev, I.~D.\ 2011, \aap, 532, A104 

\bibitem{TAvirgonew} Karachentsev, I.~D., Tully, R.~B., Wu, P.-F.,
  Shaya, E.~J., \& Dolphin, A.~E.\ 2014, \apj, 782, 4 

\bibitem{li} Li, Y.-S., \& White, S.~D.~M.\
2008, \mnras, 384, 1459 

\bibitem{Xlens} Israel, H., Reiprich, T.~H., Erben, T., et al.\ 2014, \aap, 564, A129 

\bibitem{nobound1} Kai, T., Kozaki, H., Nakao, 
K., Nambu, Y., \& Yoo, C.\ 2007, Progress of Theoretical Physics, 117,
229 

\bibitem{nobound2} Romano, A.~E.\ 2010, \prd, 82, 
123528

\bibitem{nobound3} Mattsson, T.\ 2010, General 
Relativity and Gravitation, 42, 567 

\bibitem{nobound4} Romano, A.~E, \& Chen, P.\ 2011, \jcap, 10, 16 

\bibitem{ptt} Pavlidou, V., 
Tetradis, N., \& Tomaras, T.~N.\ 2014, \jcap, 5, 17 

\bibitem{leo} Boylan-Kolchin, M. et al.\ 2013, \apj, 768, 140

\bibitem{leo2} Bellazzini, M., Gennari, N., Ferraro, F.~R., \& Sollima, A.\ 2004, MNRAS, 354, 708 

\bibitem{rines03} Rines, K., Geller, M.~J., 
Kurtz, M.~J., \& Diaferio, A.\ 2003, \aj, 126, 2152 

\bibitem[Rines et al.(2013)]{rines2013} Rines, K., Geller, M.~J., 
Diaferio, A., \& Kurtz, M.~J.\ 2013, \apj, 767, 15 

\bibitem{andrey} Gonzalez, R.~E., 
Kravtsov, A.~V., \& Gnedin, N.~Y.\ 2013, arXiv:1312.2587 

\bibitem{ketal} Karachentsev, I.~D., Dolphin, A.~E., Geisler, D., et al.\ 2002, A\&A, 383, 125 

\bibitem{fornaxmass} Drinkwater, M.~J., 
Gregg, M.~D., \& Colless, M.\ 2001, \apjl, 548, L139 

\bibitem{virgo} Urban, O., Werner, N., Simionescu, A., Allen, S.~W., B\"{o}hringer, H.\ 2011, \mnras, 414, 2101 

\bibitem{kneib} Kneib, J.-P., Hudelot, 
P., Ellis, R.~S., et al.\ 2003, \apj, 598, 804 


\bibitem{rines01} Rines, K., Geller, M.~J., 
Kurtz, M.~J., et al.\ 2001, \apjl, 561, L41 

\bibitem{shapley} Mu{\~n}oz, J.~A., \& Loeb, A.\ 2008, \mnras, 391, 1341 

\bibitem{borealis2} Batiste, M., \& Batuski, D.~J.\ 2013, \mnras, 436,
  3331 

\bibitem{borealis} Pearson, D.~W., 
Batiste, M., \& Batuski, D.~J.\ 2014, \mnras, 441, 1601 

\bibitem{coincidence} Sivanandam, N.\ 2013, \prd, 87, 083514 


















\end{thebibliography}
\end{document}